\documentclass[twocolumn,aps,prl,showpacs,floatfix,superscriptaddress,longbibliography]{revtex4-1}
\usepackage{amsfonts}
\usepackage{amssymb}
\usepackage{dcolumn}
\usepackage{graphicx}
\usepackage{bm}
\usepackage{color}
\usepackage{multirow}
\usepackage{hyperref}
\newcolumntype{d}[1]{D{.}{\cdot}{#1} }

\begin{document}

\newcommand{\NZIAS}{
Centre for Theoretical Chemistry and Physics,
The New Zealand Institute for Advanced Study,
Massey University Auckland, Private Bag 102904, 0745 Auckland, New Zealand}

\newcommand{\RUG}{
Van Swinderen Institute for Particle Physics and Gravity, University of Groningen,
Nijenborgh 4, 9747 AG Groningen, The Netherlands}

\newcommand{\FNS}{
Department of Physical and Theoretical Chemistry, Faculty of Natural Sciences, Comenius University, Mlynsk\'a dolina, 84104 Bratislava, Slovakia}

\newcommand{\UNSW}{
School of Physics, University of New South Wales, Sydney 2052, Australia}

\newcommand{\CAS}{Centre for Advanced Study at the Norwegian Academy of Science and Letters, Drammensveien 78, NO-0271 Oslo, Norway}

\title{Material Size Dependence on Fundamental Constants}

\author{Luk\'a\v{s} F. Pa\v{s}teka}
\affiliation{\FNS}
\affiliation{\NZIAS}
\affiliation{\CAS}

\author{Yongliang Hao}
\affiliation{\RUG}

\author{Anastasia Borschevsky}
\affiliation{\RUG}

\author{Victor V. Flambaum}
\affiliation{\UNSW}
\affiliation{\NZIAS}

\author{Peter Schwerdtfeger}
\affiliation{\NZIAS}
\affiliation{\CAS}

\begin{abstract}
Precise experimental setups for detection of variation of fundamental constants, scalar dark matter, or gravitational waves, such as laser interferometers, 
optical cavities 
and resonant-mass detectors, are directly linked to measuring changes in material size. Here we present calculated and experiment-derived estimates for both $\alpha$- and $\mu$-dependence of lattice constants and bond lengths of selected solid-state materials and diatomic molecules that are needed for interpretation of such experiments.
\end{abstract}


\date{\today}

\maketitle


Several unification theories and standard model (SM) extensions predict variation of fundamental constants (VFC) in space and in time \cite{Uza11,Bar05}.
It has also been hypothesized that interaction of ordinary matter with a massive scalar dark matter (DM) field can produce slow temporal drifts or oscillations in the values of the fundamental constants \cite{StaFla15,StaFla15b,ArvHua15}, while  topological defects in the dark matter field can produce transient VFC \cite{DerPos14,StaFla14,*StaFla16b}. The first transient DM detection limits were recently discussed by Wcis{\l}o et al. \cite{WciMor16}.
A possible route to observe such drifts or transient effects is through systematic measurements of transitions in atomic and molecular spectra that are sensitive to the variation of dimensionless fundamental quantities such as the fine structure constant $\alpha=e^2/\hbar c$ or the proton-to-electron mass ratio $\mu=m_p/m_e$ \cite{FlaDzu09,ChiFla09,JanBet14,PasBor15,KozLev13,SafBudDeM17}.

 Laser interferometers now reach precision far exceeding that of any spectroscopic apparatus and thus offer a new promising direction in the search for VFC \cite{StaFla16}. This line of research is directly connected to the dependence of material size on VFC, through the use of resonant-mass detectors \cite{ArvDim16,AstBab03,AstBab07,Agu11} or cryogenic sapphire and silicon oscillators \cite{TobWol10,HarNan12,WieNev16,KesHag12,WieChe14,*WieChe15}. In order to interpret such experiments, knowledge of dependence of the crystal size on the fundamental constants is needed.
  
 Theoretical investigations of size dependence of molecules and bulk materials on fundamental constants are scarce. Some studies were carried out in the context of relativistic effects and corresponding changes in periodic trends, where the dependence on the fine structure constant is considered \cite{Pyykko-1988,Sohnel-2001}. More recently, King et al. investigated the dependence of structure and bonding in small molecules on both $\alpha$ and $\mu$, with the objective of finding the hypothetical ($\alpha$, $\mu$) regimes that support biochemistry and therefore life on our planet \cite{KinSid10}. Braxmaier et al. \cite{BraPraMul01} performed an investigation of the variation of the resonance frequencies of monolithic crystal cavities with possible variation of fundamental constants through the dependence of the refractive index of the medium on $\alpha$ and $\mu$. To the best of our knowledge, no prior investigations of direct size dependence of bulk materials on fundamental constants have been carried out.

In non-relativistic physics the size of molecules and solids is proportional to the Bohr radius $a_B$. This dependence cancels out in the ratio of the sizes. The individual dependence of different compounds on the fine structure constant is determined by the difference in the relativistic effects, which are proportional to $Z^2\alpha^2$ (and higher powers of $Z^2\alpha^2$). Thus, considering the ratio of the resonance frequencies in two optical cavities made from different materials, in the non-relativistic approximation there is no dependence on $\alpha$, but such dependence appears due to relativistic corrections. 

The situation is different when we compare the resonance frequency of an optical cavity with an atomic optical frequency. For example, one measures the ratio of the Sr atomic clock frequency and the resonance frequency in a silicone cavity of length $L$ \cite{WciMor16}. Here, the ratio is proportional to  $\alpha$ already in the non-relativistic approximation since the resonator frequency depends on the speed of light $c$. Indeed, the resonator frequency is $\omega_r =c k \sim  c/\lambda \sim c/L \sim c/a_B$, atomic frequency $\omega_a \sim e^2/\hbar  a_B$, therefore $\omega_a/\omega_r \sim e^2/\hbar c=\alpha$. In this case, relativistic corrections produce additional $\alpha$ dependence.
 
Another dimensionless ratio which affects the properties of different compounds is the ratio of nuclear and electron masses; the nuclear mass is approximately proportional to the proton mass, and thus we consider the proton-to-electron mass ratio $\mu$.  
 
In this work we present  a systematic investigation of the variation of crystal lattice parameters ($a_e$ and $c_e$) and molecular bond lengths ($R_e$) due to variation of the fine structure constant and the proton-to-electron mass ratio for selected solid state and molecular systems.


\begin{table*}[ptb]
\begin{minipage}{0.82\textwidth}
\caption{Experimental and calculated bond lengths $R_e$ and their corresponding calculated fractional variation with varying fine-structure constant $\alpha$ and proton-to-electron mass ratio $\mu$.}%
\label{tab:bond}
\setlength{\tabcolsep}{6pt}
\begin{tabular}{lllllllcc}
\hline\hline
\noalign{\smallskip}
\multirow{2}{*}{mol.}	&\multirow{2}{*}{state}	&\multicolumn{3}{c}{$R_e$ [\AA]}	&\multicolumn{2}{c}{$\frac{\text{d}R_e}{R_e} / \frac{\text{d}\alpha}{\alpha}$}	& $\frac{\text{d}R_0}{R_0} / \frac{\text{d}\mu}{\mu}$ & $\frac{\text{d}R_e}{R_e} / \frac{\text{d}\mu}{\mu}$\\
\noalign{\smallskip}
		&		&\multicolumn{1}{c}{exp.\footnote{Experimental values from Refs. \cite{HubHer79,BeuKraBha93,SimHac90,HosLiClo02,BonHeaMil83,HeaMilBon83}}}	&\multicolumn{1}{c}{CC}	&\multicolumn{1}{c}{DFT}	&\multicolumn{1}{c}{CC}	&\multicolumn{1}{c}{DFT} &\multicolumn{1}{c}{Eq. (\ref{eq:dR0})} & \multicolumn{1}{c}{DBOC-CC}\\
\hline
Cu$_2$	&$^1\Sigma_g^+$	&2.2197(1)	&2.216	&2.215	&--3.19$\times10^{-2}$	&--3.18$\times10^{-2}$	              &--7.15$\times10^{-4}$ 	&--7.31$\times10^{-6}$  \\
Ag$_2$	&$^1\Sigma_g^+$	&2.5303(2)	&2.522	&2.565	&--7.66$\times10^{-2}$	&--8.21$\times10^{-2}$	              &--5.34$\times10^{-4}$ 	&--3.38$\times10^{-6}$  \\
Au$_2$	&$^1\Sigma_g^+$	&2.4719(1)	&2.471	&2.501	&--3.15$\times10^{-1}$	&--3.37$\times10^{-1}$	              &--2.95$\times10^{-4}$ 	&\hphantom{--}9.16$\times10^{-7}$  \\
C$_2$	&$^1\Sigma_g^+$	&1.24253(2)	&1.243	&1.254  	&--2.65$\times10^{-4}$ 	&--3.88$\times10^{-4}$		              &--1.17$\times10^{-3}$ 	&--6.94$\times10^{-6}$  \\
Si$_2$	&$^3\Sigma_g^-$	&2.246	&2.255	&2.309	&\hphantom{--}8.66$\times10^{-5}$ 	&\hphantom{--}1.94$\times10^{-4}$	  &--7.08$\times10^{-4}$ 	&--6.63$\times10^{-6}$  \\
Ge$_2$	&$^3\Sigma_g^-$	&2.3667(6)\footnote{For Ge$_2$, only $R_0$ was available experimentally. This was used together with the experimental $B_0$ and $\alpha_e$ value of $2.84\times10^{-4}$ cm$^{-1}$ from our CC calculations to calculate $R_e$.}	&2.361	&2.430	&--6.74$\times10^{-3}$	&--6.48$\times10^{-3}$	              &--4.37$\times10^{-4}$ 	&--4.46$\times10^{-6}$  \\
Sn$_2$	&$0_g^+$		&2.746(1)	&2.722	&2.814	&--2.24$\times10^{-2}$	&--2.17$\times10^{-2}$	                    &--3.25$\times10^{-4}$ 	&--3.82$\times10^{-6}$  \\
Pb$_2$	&$0_g^+$		&2.9271(2)	&2.869	&2.960	&--1.05$\times10^{-1}$	&--1.58$\times10^{-1}$	                    &--3.37$\times10^{-4}$   &\hphantom{--}3.32$\times10^{-7}$\\
\hline
\end{tabular}
\end{minipage}
\end{table*}

For our solid-state study we have selected several elemental and compound crystals. The choice of Cu, Si, Al, Nb and Al$_2$O$_3$ was motivated by highly precise experimental setups measuring effects of physics beyond the SM: silicon and sapphire oscillators \cite{TobWol10,HarNan12,WieNev16,KesHag12,WieChe14,*WieChe15} and resonance of Cu, Al and Nb bars \cite{ArvDim16,AstBab03,AstBab07}. In order to illustrate periodic trends we have also chosen to study the group 11 and 14 elemental solids. 

However, we initially investigated the dependence of the equilibrium interatomic distance $R_e$ of diatomic molecules on $\alpha$ and $\mu$. This gave us the opportunity to test the methodology used for the investigation of solids. 
For molecules, we have the option of using high-level \textit{ab-initio} methodology such as relativistic coupled cluster (CC) theory as a benchmark to our density functional theory (DFT) results. 
For this part of the study we chose dimers of group 11 and group 14 elements.

In order to investigate the dependence on the fine structure constant, we performed a series of optimizations of the equilibrium bond lengths $R_e$ varying the relativistic parameter $x=(\alpha /\alpha_0)^2-1$.
In case of the closed shell Cu$_2$, Ag$_2$, Au$_2$, and C$_2$
we employed the single-reference CCSD(T) method, and for the remaining open-shell systems 
(Si$_2$, Ge$_2$, Sn$_2$, Pb$_2$) the Fock space CCSD method was used. 
For the DFT part, we have used the PBE functional \cite{PBE96,*PBE97} to keep the methodology consistent with the solid-state calculations described below. Dyall's v4z basis sets \cite{Dya04,*Dya06,*Dya07} were used for all the systems except Au$_2$, where the v3z basis was used to conserve computational effort. All the calculations were carried out in the 4-component framework using the relativistic molecular program package DIRAC 15 \cite{DIRAC15}. The results are collected in Table \ref{tab:bond}.

Both CC and DFT calculated bond lengths agree with experiment. The mean absolute symmetric percentage error (MASPE) of CC results with respect to experiment is 0.5\%. Correspondingly for DFT, the MASPE is 1.6\%. Comparing the two methods to one another, we obtain the MASPE value of 1.9\%, i.e. DFT recovers the more rigorous CC results very well. 
In case of the derivative property, the $\alpha$-variation of bond lengths, there are no experimental results. However, comparing the DFT to CC results gives the MASPE of 22\%, which we use to evaluate the uncertainty on the predicted DFT $\alpha$-variation of lattice parameters presented below.

Generally, the magnitude of the $\alpha$-dependence on $R_e$ increases with increasing atomic number $Z$, as one expects. In group 11 (Cu, Ag, Au), the effect follows the well-known $\sim Z^2$ dependence as is the case with many other properties \cite{Autschbach-2002}.
For group 14 elements (C, Si, Ge, Sn, Pb), we observe a non-monotonous trend,
although the tendency of increasing $\alpha$-sensitivity magnitude for heavier elements is clearly present. This is not surprising, as we note the changes in the ground-state electronic structure in the group 14 element sequence.

The main source of the mass dependence of molecular bond lengths comes from the vibrational motion. 
Assuming that the rotational constant and bond length relationship is $B \sim R^{-2}$, and the vibrationally averaged rotational constant is
\begin{equation}
B_0=B_e-\frac{\alpha_e}{2},
\label{eq:B0}
\end{equation}
where $B_e$ is the equilibrium rotational constant and $\alpha_e$ is the vibrational-rotational coupling constant, we can express the vibrationally averaged bond length $R_0$ as
\begin{equation}
R_0=R_e\sqrt{1+\frac{\alpha_e}{2B_e-\alpha_e}}.
\label{eq:R0}
\end{equation}
Following the scaling of equilibrium constants with reduced mass $M$, $B_e \sim M^{-1}$ and $\alpha_e \sim M^{-\frac{3}{2}}$ \cite{Mul25} we arrive at the fractional variation of $R_0$ with varying $\mu$
\begin{equation}
\frac{dR_0}{R_0}=-\frac{\alpha_e}{4(2B_e-\alpha_e)}\frac{d\mu}{\mu}.
\label{eq:dR0}
\end{equation}
This simple but useful estimate can be evaluated using readily available experimental spectroscopic constants \cite{HubHer79,BeuKraBha93,SimHac90,HosLiClo02,BonHeaMil83,HeaMilBon83}. Resulting values are shown in Table \ref{tab:bond}. Due to the negative sign in (\ref{eq:dR0}) and the fact that $\alpha_e \ll B_e$ in all realistic diatomics, these are always negative. The magnitude of the effect is relatively uniform in all investigated systems. 
Following the Pekeris formula \cite{Pek34} derived for the Morse potential,
\begin{equation}\label{eq:alpha}
\alpha_e = \frac{6B_e^2}{\omega_e}\left(\sqrt{\frac{\omega_ex_e}{B_e}}-1\right),
\end{equation}
we can expect larger fractional variation of $R_0$ with $\mu$ for shallow or strongly anharmonic interatomic potentials.

A small contribution to bond length mass dependence arises from the nuclear kinetic energy terms which are neglected in the Born-Oppenheimer approximation (BOA). Within the BOA, the nuclear motion is separated from the electronic motion, and coupling terms are in practice completely neglected. As a consequence, equilibrium bond lengths are independent of the mass and this near mass-invariance is widely used in the experimental determination of bond lengths from rotational spectra of different isotopologues of the same molecule. Taking the nuclear kinetic energy term into account the $R_e$ is in fact weakly linearly dependent on the inverse reduced mass of the molecule \cite{Wat73}. This directly translates into the $\mu$-dependence $R_e \sim \mu^{-1}$. 

To estimate the non-BOA contribution to $\mu$-sensitivity of bond lengths, we included the perturbative diagonal Born-Oppenheimer corrections (DBOC) in optimizations of the investigated molecules. We employed the CCSD methodology as implemented in the program package CFOUR \cite{CFOUR,GauTaj06} together with Jorge's TZP basis set \cite{NetMun05,CamMac08,BarOli10,NetJor13}. The current implementation only allows for a non-relativistic treatment of DBOC, and the results for heavier molecules should be considered with some caution. Relativistic effects are known to increase the DBOC in atoms by up to about 40\% for the heavy elements considered in this work; however, the trends calculated for diatomic molecules in Ref. \cite{ImaAbe16} are not systematic.

Results for the fractional variation of $R_e$ are collected in Table \ref{tab:bond}. Generally, the size of this effect is comparable in all investigated diatomics. This contribution is at least two orders of magnitude smaller than the fractional variation of $R_0$ in all investigated systems and therefore it can be safely neglected when considering the $\mu$-sensitivity of vibrationally averaged bond lengths. 
One can expect this conclusion to be even more justified in solids, where the atoms are further confined in the lattice.

We estimate our $\mu$-sensitivity results to be accurate within $\pm 5\%$, considering error bars from experimental determination of spectroscopic parameters and errors introduced by neglecting the higher-order spectroscopic constants ($\gamma_e$, $\epsilon_e$ etc.) and the non-BOA effects. The error in $\alpha_e$ determination dominates the resulting compound error of the $\mu$-sensitivity.

\begin{figure}[t]
\centering
\includegraphics[scale=1]{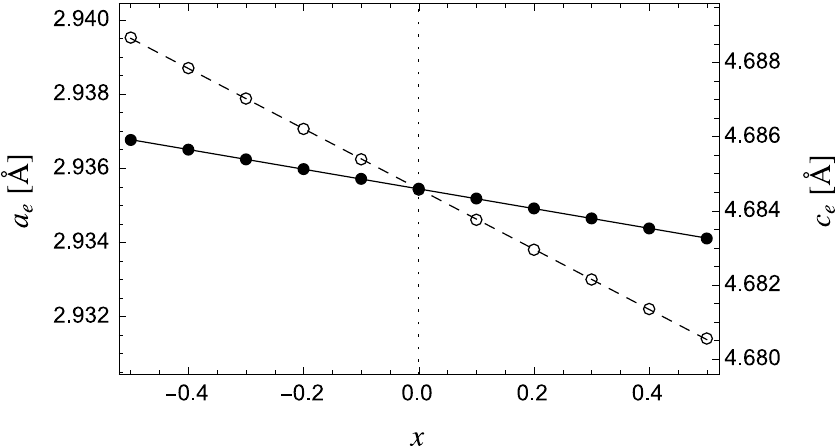}
\caption{Dependence of the lattice constants $a_e$, $c_e$ (full and open circles, 
respectively) of bulk Ti on the relativistic parameter $x$. Slopes are shown to scale.}
\label{fig:Ti}
\end{figure}

\begin{figure}[b]
\centering
\includegraphics[scale=0.7]{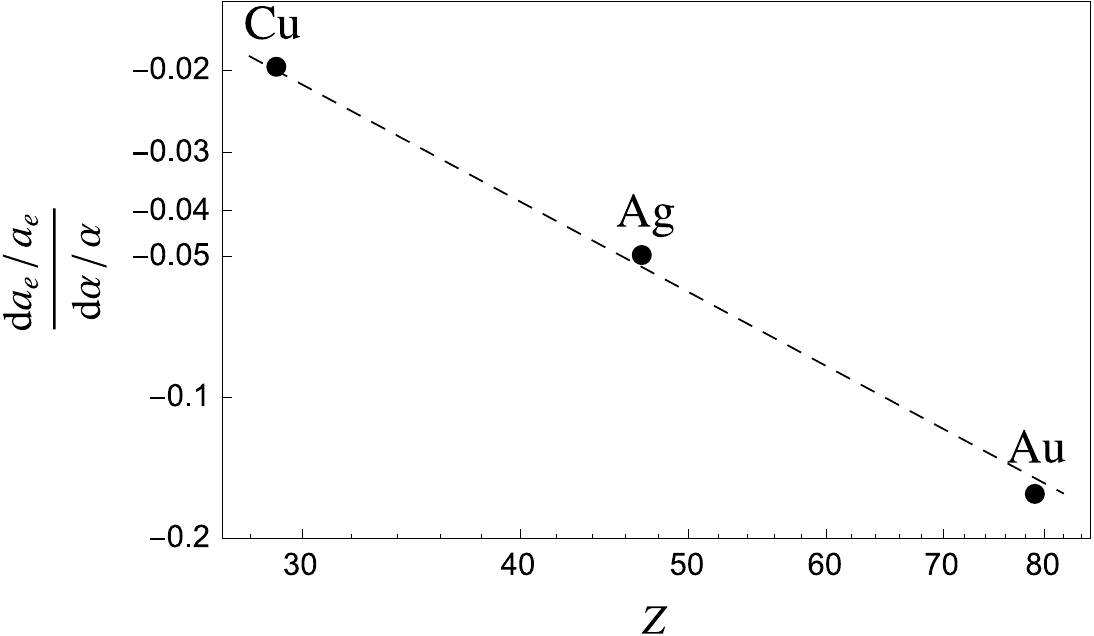}
\caption{$Z$-scaling of $\frac{da_e}{a_e} / \frac{d\alpha}{\alpha}$ for group 11 elements (note the log-log scale). Dashed line shows the ideal $Z^2$ fit.}
\label{fig:Zscaling}
\end{figure}


\begin{table*}[ptb]
\begin{minipage}{0.8\textwidth}
\caption{Experimental and calculated lattice constants $a_e$ and $c_e$ ($c_e$ shown in the second line for the respective structure) and their corresponding calculated fractional variation with varying fine-structure constant $\alpha$ and proton-to-electron mass ratio $\mu$.}%
\label{tab:lattice}
\setlength{\tabcolsep}{6pt}
\begin{tabular}{lllrrrcc}
\hline\hline
\noalign{\smallskip}
\multirow{2}{*}{solid}	&\multirow{2}{*}{structure}	&\multirow{2}{*}{spc. group}		&\multicolumn{3}{c}{$a_e,\,c_e$ [\AA]}	&\multirow{2}{*}{$\frac{\text{d}a_e}{a_e} / \frac{\text{d}\alpha}{\alpha}$}	&\multirow{2}{*}{$\frac{\text{d}a_0}{a_0} / \frac{\text{d}\mu}{\mu}$}\\
\noalign{\smallskip}
		&			&		&exp. (RT)\footnote{Experimental values from Refs. \cite{CRC,MasStrStr93,LanBol41A1b,LitShaYin10}}& exp. (0K)\footnote{Finite-temperature experimental lattice parameters extrapolated to 0K \cite{MitGir86} using experimental data from Refs. \cite{Kit05,Wal02,AleGonZis87,Sti00,Gon87,LanBol41A1a,LanBol41A1b,RayCha60,AltBakTru62,TarLedOgi13,LitShaYin10}}	&calc.	&\\
\hline
Cu		&fcc			&$Fm\bar{3}m$	&3.6146(2)	&3.6029(2)	&3.634	&--1.97$\times10^{-2}$		&--1.12$\times10^{-3}$\\
Ag		&fcc			&$Fm\bar{3}m$	&4.0857(2)	&4.0681(2)	&4.160	&--4.97$\times10^{-2}$		&--8.32$\times10^{-4}$\\
Au		&fcc			&$Fm\bar{3}m$	&4.0782(2)	&4.0646(2)	&4.059	&--1.61$\times10^{-1}$		&--4.33$\times10^{-4}$\\

C		&dia.			&$Fd\bar{3}m$	&3.5669(2)	&3.5667(2)	&3.576	&--2.39$\times10^{-4}$		&--2.22$\times10^{-3}$\\
Si		&dia.			&$Fd\bar{3}m$	&5.4306(2)	&5.4259(2)	&5.479	&--2.17$\times10^{-4}$		&--8.56$\times10^{-4}$\\
Ge		&dia.			&$Fd\bar{3}m$	&5.6574(2)	&5.6487(2)	&5.779	&\hphantom{--}6.21$\times10^{-4}$&--6.02$\times10^{-4}$\\
Sn		&dia. ($\alpha$)&$Fd\bar{3}m$	&6.4892(2)	&6.4752(2)	&6.678	&--4.46$\times10^{-4}$		&--3.29$\times10^{-4}$\\
		&tet.	($\beta$)	&$I4_1/amd$	&5.8318(2)	&5.8048(2)	&5.956	&\hphantom{--}7.61$\times10^{-3}$&--7.09$\times10^{-4}$\\
		&			&		           	&3.1818(2)	&3.1671(2)	&3.251	&\hphantom{--}1.16$\times10^{-3}$&\\
Pb		&fcc			&$Fm\bar{3}m$	&4.9502(2)	&4.9142(2)	&4.715	&--2.94$\times10^{-1}$		&--5.65$\times10^{-4}$\\					

Al		&fcc			&$Fm\bar{3}m$	&4.0496(2)	&4.0321(2)	&4.042	&\hphantom{--}3.65$\times10^{-4}$&--2.11$\times10^{-3}$\\
Nb		&bcc			&$Im\bar{3}m$	&3.3004(2)	&3.2955(2)	&3.317	&--2.35$\times10^{-3}$		&--3.74$\times10^{-4}$\\
Ti		&hcp			&$P6_3/mmc$	    &2.9506(2)	&2.9461(2)	&2.935	&--1.18$\times10^{-3}$		&--7.06$\times10^{-4}$\\
		&			&		        	&4.6835(2)	&4.6764(2)	&4.685	&--3.47$\times10^{-3}$		&\\
Al$_2$O$_3$&hex.		&$R\bar{3}c$	&4.7540(5)	&4.7507(5)	&4.825	&--4.06$\times10^{-4}$		&--2.06$\times10^{-3}$\\
		&			&		        	&12.9820(6)	&12.9731(6)	&13.142	&--4.88$\times10^{-4}$		&\\
SiC		&3C ($\beta$)	&$F\bar{4}3m$	&4.3596(1)	&4.3582(1)	&4.392	&--3.54$\times10^{-4}$		&--1.56$\times10^{-3}$\\
		&6H ($\alpha$)	&$P6_3mc$	    &3.0806(1)	&3.0795(1)	&3.105	&--3.73$\times10^{-4}$		&--1.47$\times10^{-3}$\\
		&			&			        &15.1173(1)	&15.1121(1)	&15.225	&--3.41$\times10^{-4}$		&\\
WC		&hex.		&$P\bar{6}m2$	    &2.9059(1)	&2.9051(1)	&2.923	&--4.39$\times10^{-2}$		&--1.26$\times10^{-3}$\\
		&			&			        &2.8377(1)	&2.8369(1)	&2.857	&--3.19$\times10^{-2}$		&\\
\hline
\end{tabular}
\end{minipage}
\end{table*}


The investigation of the $\alpha$ and $\mu$-sensitivity of lattice parameters of solids was conceptually analogous to the initial study on diatomics.
For the $\alpha$-dependence, series of optimizations of the $a_e$ and $c_e$ equilibrium lattice parameters were performed varying the relativistic parameter $x$. Technically, this means the value of the speed of light $c$ was varied in the computations. All calculations were carried out using the relativistic DFT solid-state program package FPLO 14.0, which implements the full-potential local-orbital minimal-basis scheme \cite{KoeEsc99,EscRic04}. We tested several DFT functionals and decided to use the gradient corrected PBE functional \cite{PBE96,*PBE97}, which had the smallest error in the equilibrium lattice parameters compared to the available experimental values. In all systems, reciprocal sampling used a $k$-mesh of $12 \times 12 \times 12$ and 200-point radial mesh. Both were saturated with respect to the equlibrium lattice constants and their derivatives.
Optimizations of the lattice constants were performed within a fully relativistic framework, while the internal parameters (atomic positions within the unit cell) were optimized using scalar relativity in each step due to the limitation of the current implementation (the latter only applies to Al$_2$O$_3$ and $\alpha$-SiC).
Optimal lattice parameters were determined from minima of polynomial fits through grids of calculated points with nummerical accuracy of $10^{-7}$ \AA.
Additionally, we compared the results of our FPLO calculations on Cu, C (diamond), and Si to those obtained from an independent program package SPRKKR 6.3 \cite{SPRKKR6.3, EbeKod11} based on the spin polarized relativistic Korringa-Kohn-Rostoker methodology \cite{Ebe00}. The agreement between the two data sets was excellent with discrepancies below 2\%.

The results of our DFT solid-state calculations are collected in Table \ref{tab:lattice}. The calculated equilibrium lattice parameters are in a very good agreement with experimental values, giving the MASPE value of 1.3\%, which is slightly smaller than in the case of diatomic molecules. Relying on the error analysis of diatomics, we thus expect the calculated $\alpha$-dependence of lattice constants to be accurate within $\pm 20\%$. Figure \ref{fig:Ti} demonstrates that the lattice parameters exhibit close to ideal $\alpha^2$ scaling, i.e. their dependence on $x$ is almost perfectly linear across a wide range of $\alpha$ values (in the present study, $\sqrt{0.5}\text{ }\alpha_0 \leq \alpha \leq \sqrt{1.5}\text{ }\alpha_0$).
We note that the values of fractional variation $\frac{da_e}{a_e} / \frac{d\alpha}{\alpha}$ are relatively small ($\ll1$).

We once more observe the general increase of the $\alpha$-sensitivity with $Z$. In the case of group 11 elements, this increase follows the $\sim Z^2$ scaling (Figure \ref{fig:Zscaling}). For group 14, the overall effect is highly non-monotonous. This is also true if we only compare crystal structures with the same space group $Fd\bar{3}m$ (C, Si, Ge, $\alpha$-Sn), as the bonding character changes significantly along this sequence from strongly covalent to semi-metallic \cite{Hermann-2010}. Another interesting feature is that the $\alpha$-sensitivity of diamond, the lightest element in the group, is larger than expected from the simple scaling law, and larger than that of its heavier homologue, Si.
Comparing the two allotropes of tin, we observe opposite sign of the size dependence on $\alpha$. 
Furthermore, the absolute $\alpha$-sensitivity of $\beta$-Sn is higher than that of $\alpha$-Sn. This can be ascribed to relativistic effects more strongly influencing the distorted-close-pack structure of $\beta$-Sn, which has higher s character compared to fully sp$^3$-hybridized diamond structure of $\alpha$-Sn \cite{LegMan16}. Relativistic effects are strongest in s orbitals with the highest density in the vicinity of the nucleus. Population analysis gives 5s:1.54$e$, 5p:2.34$e$ for $\alpha$-Sn and 5s:1.73$e$, 5p:2.14$e$ for $\beta$-Sn, supporting this interpretation.
For the non-cubic crystal structures (corundum, $\beta$-tin, $\alpha$-SiC, WC) we see different $\alpha$-dependence of the lattice constants $a_e$ and $c_e$; especially in the case of $\beta$-Sn, where the $\alpha$-sensitivity of $a_e$ is roughly $6\times$ larger than that of $c_e$. Thus, the character of bonding has a very strong influence on the dependence of the crystal structure parameters on $\alpha$, and in the general case scaling laws are not sufficient to estimate the size of the effect.

A general practice for removing the dependence on the unit system
is to use the ratio of two observed quantities instead of measuring a property of a single system. In this case, we can either compare two different materials (preferably with opposite signs of their $\alpha$-dependencies, to enhance the sensitivity) or even compare the $\alpha$-sensitivity in two different directions of a single material (for non-cubic crystals). Additionally, in interferometry, the observed quantity is the phase shift, which is unitless in itself, and one type of measurement would suffice for topological DM detection.

The mass dependence of lattice parameters is due to the vibrational motion of the crystal lattice. It was extensively studied theoretically and experimentally using isotopic substitution.
For monoatomic solids at zero temperature (where the mass dependence is largest \cite{PavBar94}), London derived an expression for fractional variation of molar volume $V$ with varying isotopic molar mass $M$ \cite{Lon58} 
\begin{equation}
\frac{MdV}{VdM}=-\frac{9}{16}\frac{\gamma\kappa}{V}R\Theta_D,
\label{eq:London}
\end{equation}
where $\gamma$ is the thermodynamic Gr\"{u}neisen parameter, $\kappa$ the compressibility, $R$ the gas constant, and $\Theta_D$ the Debye temperature. 
This translates to a final expression for the $\mu$-variation of the lattice constant
\begin{equation}
\frac{da_0}{a_0}/\frac{d\mu}{\mu}=-\frac{3}{16}\frac{\gamma}{B V}k_B\Theta_D,
\label{eq:mu-var}
\end{equation}
where $B$ is the isothermal bulk modulus, $V$ is atomic volume and $k_B$ is Boltzmann constant. Since all numbers entering this formula are positive, the resulting values are negative for all realistic crystals. Note, the bulk modulus and hence also the $\mu$-variation are assumed to be isotropic.

Results obtained using expression (\ref{eq:mu-var}) and available experimental parameters \cite{Kit05,Wal02,AleGonZis87,Sti00,Gon87,LanBol41A1a,LanBol41A1b,RayCha60,AltBakTru62,TarLedOgi13,LitShaYin10}
are listed in Table \ref{tab:lattice}. 
Considering the errors in experimental determination of solid-state parameters, we estimate the overall presented $\mu$-sensitivity values to be accurate within $\pm 20\%$.
Our $\mu$-sensitivity estimates compare well with the results derived from available experimental measurements on diamond, silicon and germanium (--1.8$\times10^{-3}$, --7.8$\times10^{-4}$ and --5.5$\times10^{-4}$, respectively) \cite{HolHas91, *HolHas92, WilShv02, HuSin03}, as well as with results derived from theoretical path-integral Monte Carlo simulations (C: --1.94$\times10^{-3}$; Si: --1.13$\times10^{-3}$; Ge: --6.55$\times10^{-4}$; $\beta$-SiC: --1.37$\times10^{-3}$) \cite{Her01,Her99,NoyHer97,HerRam09}. The MASPE value of our estimates with respect to the referenced results is 15\%, supporting our error analysis.
Out of the investigated materials, diamond, Al and Al$_2$O$_3$ (i.e. corundum or sapphire) have the highest $\mu$-sensitivity. All results lie in a relatively small range spanning one order of magnitude. Within this range, however, there are no clear systematic trends.

Precision interferometry can now provide relative sensitivity of parameters such as $\delta L/L$ to VFC beyond that of any other physical apparatus. Combining this with the calculated fractional variation of crystal sizes and the use of silicon, sapphire or other single crystal oscillators or optical cavities offers a new path to testing VFC and search for scalar low-mass dark matter beyond the most stringent limits \cite{StaFla16}. For a progress in this field, accurate values for $\alpha$- and $\mu$-dependence of lattice constants of solid-state materials are required, which we provide in this work.

The authors are grateful to the Mainz Institute for Theoretical Physics (MITP) for its hospitality
and its partial support during the completion of this work.



%

\end{document}